\documentclass{ptephy_v1}

\usepackage{amsmath}
\usepackage{amssymb}

\usepackage{graphicx}
\usepackage{dcolumn}
\usepackage{bm}
\usepackage{color}

\usepackage[dvipdfmx,
colorlinks=true,
urlcolor=blue,
citecolor=blue,
linkcolor=blue,
hyperfootnotes=false]{hyperref}

\Year{2024}

\begin{document}

\title{Holographic analysis of boundary correlation functions for the hyperbolic-lattice Ising model
}

\author{Kouichi Okunishi${}^{1}$}
\author{Tomotoshi Nishino${}^{2}$}

\affil{${}^1$Department of Physics, Niigata University,  Niigata 950-2181, Japan}
\affil{${}^2$Department of Physics, Graduate School of Sciences, Kobe University, Kobe 657-8501,Japan}
\date{\today}

\begin{abstract}
We analyze boundary spin correlation functions of the hyperbolic-lattice Ising model from the holographic point of view. 
Using the corner-transfer-matrix renormalization group (CTMRG) method, we demonstrate that the boundary correlation function exhibits power-law decay with quasi-periodic oscillation, while the bulk correlation function always decays exponentially. 
On the basis of the geometric relation between the bulk correlation path and distance along the outer edge boundary, we find that scaling dimensions for the boundary correlation function can be well explained by the combination of the bulk correlation length and background curvatures inherent to the hyperbolic lattice.
We also investigate the cutoff effect of the bond dimension in CTMRG, revealing that the long-distance behavior of the boundary spin correlation is accurately described even with a small bond dimension.
In contrast, the sort-distance behavior rapidly loses its accuracy. 
\end{abstract}

\maketitle


\section{ Introduction \label{tag_sec1}}

The tensor network has been rapidly developed as an essential theoretical framework for efficiently describing quantum/classical many-body systems in various physics research fields.\cite{JPSJ2022, Orus2019}
The crucial point is that a network integration of tensors representing``propagations" of quantum entanglements is realized through renormalization group (RG) like coarse-graining of physical degrees of freedom.
In particular, the multi-scale entanglement renormalization ansatz (MERA)\cite{MERA2007} for one-dimensional(1D) quantum systems and tensor network renormalizations \cite{TNR2015,Evenbly2015, LoopTNR2017, GILT2018, Harada2018} for 2D classical models provide scale-invariant tensor network states capable of representing power-law correlation functions and the area-law\cite{Eisert_RMP2010} of entanglement entropy up to log-correction.
Moreover, the intrinsic connection of such scale-invariant tensor networks with AdS/CFT\cite{Maldacena1998,Gubser1998,Witten1998,Aharony2000} and the Ryu-Takayanagi formula for the holographic entanglement entropy \cite{RT_PRL2006, RT_JHEP2006} has elucidated the necessity of a more comprehensive investigation about the scale invariance mechanism behind the tensor network structure associated with the holography principle.\cite{Swingle2012,Matsueda2012,Happy,Hyden2016,Evenbly2017}

In order to gain a deeper understanding of the holography principle behind such scale-invariant tensor networks, an essential step is to investigate a simple but nontrivial spin model that mimics the hyperbolic geometry relevant to the AdS space-time. 
At the level of the tree tensor network,  recently,   the precise relationship of the Bethe lattice Ising model with the holographic RG and the p-adic version of AdS/CFT\cite{Gubser2017,Heydeman2016,Bhattacharyya2017,Chen2021,Yan2023} was revealed, where the hyperbolic network geometry plays a crucial role in ensuring the correspondence of scaling dimensions for a boundary spin operator and its conjugate magnetic field.\cite{Oku2023}
Nevertheless, the Bethe lattice model, which is a typical case of the tree tensor network containing no loop structure in its network geometry, does not satisfy log correction to the area law of the entanglement entropy.

\begin{figure}[tb]
\begin{center}
\includegraphics[width=6.5cm]{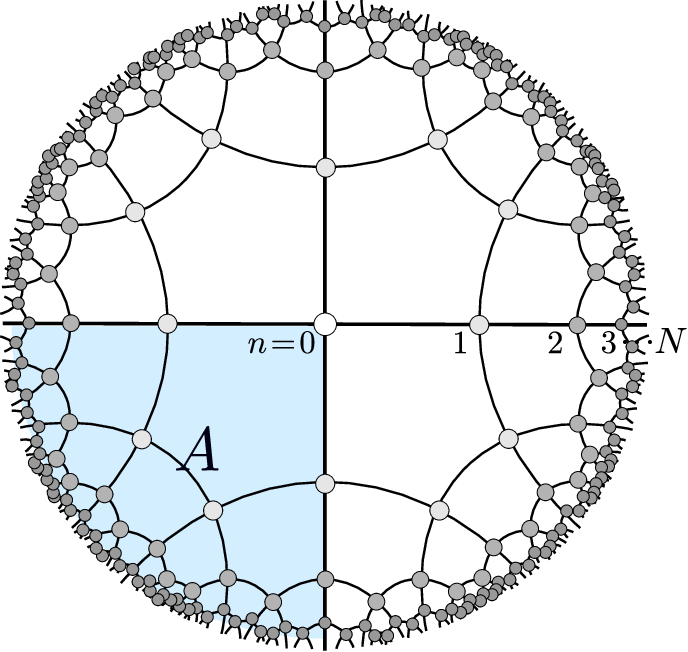}~~
\end{center}
\caption{A $\{5,4\}$ hyperbolic-lattice Ising model illustrated on the Poincar\'e unit disk, where all regular pentagons are equivalent in the hyperbolic space. 
The circles on the lattice nodes represent Ising spins; Nodes near the outer edge of the unit disk are omitted for simplicity.
The generation index $n$ for the nodes along the horizontal geodesic line corresponds to the hyperbolic distance from the center node, and $N$ denotes the generation index for the outer edge node.
One of the four quadrants sharing the center node (shaded region) corresponds to a CTM $A$.
}
\label{fig1}
\end{figure}

In the above context,  the lattice discretitzation of field theories based on hyperbolic lattices provides an interesting lattice counterpart of physics on the AdS side.\cite{Boettcher2020,Asaduzzaman2020,Brower2021,Basteiro2023BF} 
Moreover, the hyperbolic-lattice Ising model can be a fundamental statistical mechanics model, which has attracted much attention from the viewpoints of tensor networks and AdS/CFT.\cite{Shima2006,UedaK2007,Krcmar2008,Lee2016,Asaduzzaman2022}
An intrinsic feature of the hyperbolic tiling lattice,  which is often viewed as a lattice version of Euclidean AdS$_2$, is that the loop structure due to polygons is compatible with lattice translation symmetry.
Recent numerical analyses of such hyperbolic-lattice models revealed that the bulk spin correlation function decays exponentially\cite{Iharagi2010,Gendiar2012}, while intriguing power-law decay of correlation functions was observed for boundary spins\cite{Asaduzzaman2022}.
A lattice translation along the geodesic line on the hyperbolic-lattice network basically corresponds to a shift of scale layers in the MERA-type tensor network.
Thus, a precise tensor network analysis for boundary spin correlation functions and entanglements of the hyperbolic-lattice model is indispensable for elucidating the relation of network geometry and holographic RG structure behind scale-invariant tensor networks.

In this paper, we investigate the boundary spin correlation function for the Ising model defined on a hyperbolic lattice using the corner-transfer-matrix renormalization group (CTMRG) technique\cite{CTMRG1}.
The corner transfer matrix (CTM) was initially formulated as a variational state for the row-to-row transfer matrix of the planar lattice based on the matrix-product-state representation.\cite{Baxter1968,Baxterbook,JPSJ2022}
For the hyperbolic-lattice model, the number of spins in the row-to-row transfer matrix grows exponentially as the system size increases, where a matrix product state decomposition of its eigenvector becomes nontrivial.\cite{Oshima2024} 
However, the variational principle based on the reduced density matrix in CTMRG works efficiently\cite{CTMRG2} and enables us to obtain the bulk expectation value of thermodynamic quantities for hyperbolic-lattice systems.\cite{Iharagi2010,Gendiar2012}
For a $\{5,4\}$ hyperbolic-lattice Ising model, we demonstrate the power-law decay of boundary spin correlation functions with quasi-periodic oscillation and discuss their connection to hyperbolic geometries.
Here, $\{p,q\}$ denotes the Schl\"{a}fli symbol.
Moreover, we analyze the bond-dimension dependence of the boundary spin correlation function with CTMRG.
We then find that the cutoff effect becomes significant on its short-distance behavior rather than the long-distance one, which is reminiscent of the Kadanoff-Wilson type RG where  the infra-red behavior can be selectively extracted by tracing out ultra-violet (UV) degrees of freedom\cite{WilsonRG,Kadanoff2014}.
We also discuss how the bond-dimension dependence of the boundary correlation function is modified from the planar lattice case, refering to the spectrum of the reduced density matrix in CTMRG calculations.

This paper is organized as follows.
In Sec. 2, we briefly summarize the hyperbolic-lattice Ising model and details of the CTMRG method.
In Sec. 3, we calculate boundary spin correlation functions and demonstrate the power-law decay, referring to the bulk correlation path along geodesics.
In Sec. 4, we discuss the crossover behavior of the boundary spin correlation functions, originating from the bulk order in the low-temperature phase.
In Sec. 5, we investigate the cutoff dependence of boundary spin correlations in CTMRG computations. 
We then show that the longer-range correlation is maintained even with a small bond dimension, but the short-distance correlation is fragile against the cutoff effect.
Sec. 6 is devoted to the summary and discussions.

\section{The hyperbolic-lattice Ising model and CTMRG }

We consider the Ising model on a $\{5, 4\}$ lattice as a typical example of the hyperbolic-lattice model, which is depicted in Fig. \ref{fig1}.
In this paper, we define the edge length of each regular pentagon as a unit-length scale.
For the lattice geometry in Fig. \ref{fig1}, then, we assign the generation index $n$ for nodes from the center ($n=0$) to the outer edge ($n=N$) along the horizontal (or vertical) line, which corresponds to the hyperbolic distance\footnote{In this paper, we call the distance along a geodesic line in the background hyperbolic plane ``hyperbolic distance'' for short.} of the node from the center.
In addition, we define the system size as the generation index $N$ for the outer edge node.

\begin{figure}[bt]
\begin{center}
\includegraphics[width=10cm]{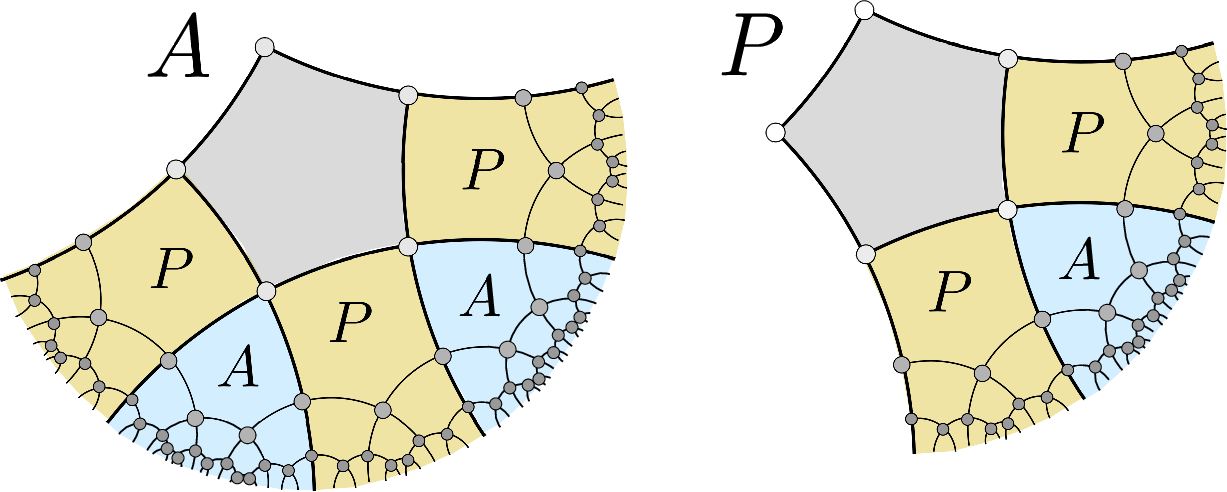}~~
\end{center}
\caption{
Recursive relations for the CTM $A$ and  half of the row-to-row transfer matrix $P$ in CTMRG. 
}
\label{fig2}
\end{figure}

\subsection{$\{5,4\}$ lattice model and CTMRG}

\begin{figure}[tb]
\begin{center}
\includegraphics[width=6cm]{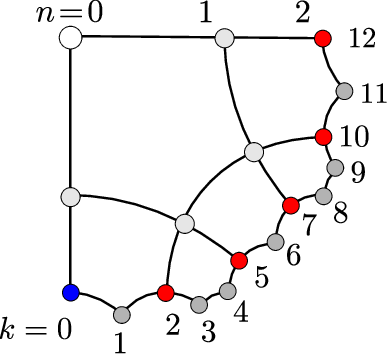}~~
\end{center}
\caption{A CTM with $N=2$ for the $\{5,4\}$ hyperbolic lattice.
The index $k$ represents the location of boundary nodes in the circumference direction. 
We calculate the spin correlation function $\langle\sigma_{N,0}\sigma_{N,k}\rangle$ for the spin at $k=0$ (blue circle) and spins at $k=2$, 5, 7, 12 (red circles).}
\label{fig3}
\end{figure}

The partition function of the model is explicitly written as 
\begin{align}
\mathcal{Z} = \sum_{\{\sigma \}} \exp\left( K \sum_{\langle i, j \rangle} \sigma_i \sigma_j - h_b \sum_{i \in \mathrm{edge}} \mu_i \sigma_i  \right) \, ,
\end{align}
where $\sigma_i = \pm 1$ denotes an Ising spin variable at the $i$th node on the $\{5,4\}$ lattice, and $\langle i, j \rangle$ represents the sum with respect to the nearest neighboring pairs.
$K\equiv\beta J$ denotes an inverse temperature with the exchange coupling $J=1$.
Also, $h_b$ represents a magnetic field applied to spins at the outer edge boundary, and thus, no magnetic field exists in the bulk region. 
Moreover, we have introduced an effective coordination number for edge spins as $\mu= 1$ or $2$, taking into account the number of bonds connected with edge spins;
This is because, if we think of a holographic renormalization of spins from the outer side, the resulting boundary magnetic fields depend chiefly on the number of bonds linked to the outer spins traced out.
In Fig. \ref{fig3}, for example, $\mu=1$ for $k=0$, 2, 5, 7, 10, and 12,  while $\mu=2$ for $k=1$, 3, 4, 6, 8, 9, and 11. 
We assume the free boundary condition for the outer edge boundary spins.

The CTMRG for the hyperbolic-lattice model consists of recursive relations for the CTM $A$ and the hyperbolic-lattice counterpart of half of the row-to-row transfer matrix $P$\cite{UedaK2007,Krcmar2008}, which can be depicted in Fig. \ref{fig2}.  
Note that although the shape of $A$ in Fig. \ref{fig2} looks distorted compared with the quadrant in Fig. \ref{fig1}, it represents a quadrant of hyperbolic lattices with different system sizes.
Also, as in Fig. \ref{fig2}, $P$ is always sandwiched by two adjacent parallel lines in the hyperbolic lattice, whose width rapidly extends toward the outer edge boundary. 
Accordingly, the number of spins in $P$ and $A$ increases exponentially toward the outer edge boundary, as illustrated in Table \ref{table1}.
This point is one of the fundamental differences from a planar (flat) lattice system, where the size of $P$ increases in proportion to the linear dimension of the system.
However, the recursive relations in Fig. \ref{fig2} enable us to increase the size of both $P$ and $A$ iteratively, starting from appropriate initial (boundary) tensors.
After a sufficient number of iterations ($N \gg 1$), we can accurately compute the bulk expectation value of a local magnetization and bond energy at the center of the system.
For instance, the bulk transition point for the $\{5,4\}$ lattice is precisely evaluated as $K_c = 0.357$($T_c=2.799$).\cite{UedaK2007}

In this paper, we focus correlation functions for boundary spins under the geometry of Fig. \ref{fig1}, for which the partition function is explicitly written as 
\begin{align}
\mathcal{Z}=\sum_{\{\sigma \}} A^4\, .
\label{eq_z_for_fig1}
\end{align}
For Fig. \ref{fig1},  the hyperbolic distance from the center node to the outer edge boundary along the cutting edge lines of a CTM corresponds to the system size $N$.
Thus, the situation of Eq. (\ref{eq_z_for_fig1}) is convenient in discussing the relation of the resulting correlation function with the hyperbolic geometry.

As in Fig. \ref{fig3}, we introduce the index $k$ representing the location of boundary nodes along the circumference of a CTM with the system size $N$. 
We then write the boundary spin with the generation index $N$ located at the node index $k$ as $\sigma_{N,k}$ and implement a calculation of $\langle\sigma_{N,0}\sigma_{N,k}\rangle$ using the CTMRG for Eq. (\ref{eq_z_for_fig1}).
For this purpose, it is necessary to trace how the boundary pentagon with an edge spin operator is embedded in the CTMs during CTMRG iterations.
By introducing a sequence of pentagons aligned from the center to the outer edge, we can systematically construct a renormalized representation of the CTM with a spin operator at its outer edge.
After performing CTMRG iterations up to the $N$th generation, we reconstruct the renormalized CTM with edge spin operator at every $k$ and then obtain $\langle\sigma_{N,0}\sigma_{N,k}\rangle$.
Also, we calculate the correlation function between the edge spin at $k=0$ and an interior spin on the geodesic line passing through the center spin within the standard CTMRG framework.

\begin{table}
\begin{center}
\begin{tabular}{|c|r|r|}
\hline 
\hspace{1mm}$N$ \hspace{1mm} & \hspace{1cm} $\mathcal{N}_A$ \hspace{2mm} & \hspace{1cm} $\mathcal{N}_P$ \hspace{2mm} \cr
\hline
1 & 4 & 3 \cr
2 & 13 & 8 \cr
3 & 46 & 27 \cr
4 & 169 & 98 \cr
5 & 628 & 363 \cr
6 & 2341 & 1352 \cr
7 & 8734 & 5043 \cr
8 & 32593 & 18818 \cr
9 & 121636 & 70227 \cr
10 & 453949 & 262088 \cr
\hline
\end{tabular}
\end{center}
\caption{The numbers of boundary spins $\mathcal{N}_A$ and $\mathcal{N}_P$ respectively contained in $A$ and $P$ at the $N$th generation.}
\label{table1}
\end{table}

\subsection{geometric properties of the hyperbolic lattice.}
\label{subsec_geometry}

As in Figs. \ref{fig1}, the hyperbolic lattice is usually drawn on Poincar\'e unit disk.
Here, we briefly summarize the geometric property of the hyperbolic tiling with a regular polygon whose edge length is normalized to be unity.  
As in Appendix \ref{appendix_a}, the hyperbolic distance for the Poincar\'e unit disk coordinate is generally defined by the metric (\ref{eq_poincare_unitdisk}) with the Poincar\'e radius $L$. 
For the polygon tiling on the hyperbolic plane, a crucial point is that the shape of a polygon defines a particular value of $L$;
Let us write the radius for the $\{p,q\}$ tiling as $L_{pq}$.
Then, the hyperbolic version of the cosine formula provides 
\begin{align}
L_{pq}= \frac{1}{2\cosh^{-1}\left( \frac{\cos(\pi/p)}{\sin(\pi/q)}\right) }\, ,
\end{align}
implying that the curvature of the background hyperbolic plane depends only on $p$ and $q$.\cite{Mosseri1982,Coxeter1965}
Here, note that  
\begin{align}
L_{54}=0.9422\cdots\, .
\label{eq_L54}
\end{align}
for the $\{5, 4\}$ tiling.
At Poincar\'e unit disk level, an approximated radius $r_n$ associated with the $n$th generation circle is given by
\begin{align}
r_n = \tanh(\frac{n}{2 L_{pq}}) \, ,
\label{eq_rn_pq}
\end{align}
which corresponds to the location of nodes with the index $n=0, 1, \cdots$ in Fig. \ref{fig1}.[See Eq. (\ref{eq_poincare_rn})]

In turn, we can also introduce another effective coordinate for the hyperbolic tiling lattice based on the number of nodes along its outer edge circle, different from the above standard definition of $L_{pq}$. 
Let us consider lattice nodes with a generation index $n$, which are approximately distributed on a circle defined by a hyperbolic distance $n$ from the center of the disk.  
As in Appendix \ref{appendix_a}, in general, the circumference of the circle with a hyperbolic distance $\rho$ in the hyperbolic plane with a Poincar\'e radius $L$ is given by $2\pi L \sinh \left( \frac{\rho}{L} \right) \sim \exp\left( \frac{\rho}{L} \right)$ for $\rho \gg L$ [See Eq. (\ref{eq_hbcircum})].
Assuming a uniform distribution of the nodes with the $n$th generation in the circumference direction, we may approximately write the $n$ dependence of the number of lattice nodes as
\begin{align}
 \mathcal{N}_b \sim \exp\left( \frac{n}{L} \right) \, .
\label{eq_bnumL}
\end{align}
Meanwhile, we can numerically estimate $\mathcal{N}_b$ for the $\{5,4\}$ lattice in Fig. \ref{fig1} as
\begin{align}
\mathcal{N}_\mathrm{b} \propto \mathcal{N}_A \sim {p_\mathrm{eff}}^n \, 
\label{eq_bnump}
\end{align}
with $p_\mathrm{eff} = 3.73\cdots$ from Table 1.\cite{Serina2016}
By comparing Eqs. (\ref{eq_bnumL}) and (\ref{eq_bnump}), we can estimate the effective Poincar\'e radius as  
\begin{align}
L_\mathrm{eff}\equiv  1/\log p_\mathrm{eff} = 0.7598\cdots\, ,
\label{eq_Leff}
\end{align}
which clearly deviates from Eq. (\ref{eq_L54}).
As in the case of the Bethe lattice, then, we may describe the distribution of nodes by introducing an effective coordinate corresponding to Poincar\'e upper half (\ref{eq_upperhalf}) with compactification,  summarized in Appendix \ref{appendix_a2}.
\footnote{For the Bethe lattice case in Ref. \cite{Oku2023}, all the edge nodes of the tree network were naively arranged in the outer circle of the disk-like representation, for which the effective metric near the outer edge boundary corresponds to (compactified) Poincar\'e upper-half plane, although the appearance of the tree network looks as if Poincar\'e unit disk.}
Here, it may be interesting to note that $L_\mathrm{eff}$ is based on the macroscopic property (\ref{eq_bnumL}) for the hyperbolic plane.
In contrast, $L_{pq}$ purely originates from the microscopic structure of the regular polygon tiling.

In the next section, we will see that both Eqs. (\ref{eq_L54}) and (\ref{eq_Leff}) play fundamental roles in analyzing the boundary spin correlation functions.

\section{Correlation functions for the $\{5,4\}$ lattice model}

We calculate exact spin correlation functions for the $\{5,4\}$ lattice model up to $N=10$ in the setup of Fig. \ref{fig1}, using CTMRG with no cutoff of the bond dimension ($\chi = 1024$ for $N=10$).
We first discuss the bulk properties along the geodesic line through the center node of the lattice and then analyze outer edge spin correlation functions in connection with disk-like coordinates associated with the Euclidean AdS$_2$.

\subsection{spin correlations along the geodesic line}
\label{sec_3A}

We investigate the property of spin correlation functions along the geodesic line for the free boundary system.
For this purpose, as in Fig. 1, let us consider the horizontal geodesic line passing through the center node, where we write a spin at the generation index $n$ as $\sigma_n$ with omitting the circumference-direction index for simplicity.
We then calculate $\langle \sigma_{N} \sigma_{n} \rangle$ for $N=10$ with CTMRG, where $\sigma_N$ is fixed at the outer edge of the left side block and $\sigma_n$ moves along the horizontal line toward its antinodal edge in the right.  
Note that the hyperbolic distance between two spins along the geodesic line is explicitly written as
\begin{align}
\tilde{\rho} \equiv \begin{cases}
 N-n \quad  \quad \text{left side block} \\
 N+n \quad  \quad \text{right side block}
 \end{cases} \, .
 \label{eq_def_rho}
 \end{align}

\begin{figure}[bt]
\begin{center}
\includegraphics[width=6cm]{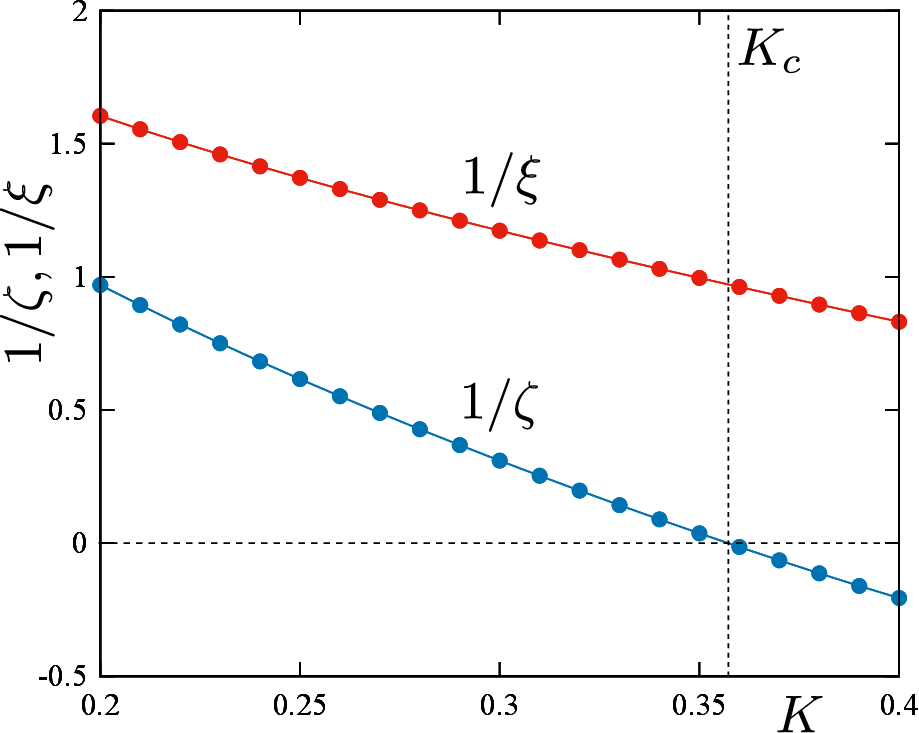}~~
\end{center}
\caption{The inverse correlation length $1/\xi$ along the geodesic line for $h_b=0$ and the inverse penetration depth $1/\zeta$ of the local magnetization under a weak boundary magnetic field($h_b=0.01$).}
\label{fig4}
\end{figure}

An important point is that the correlation function always exhibits exponential decay even around the bulk transition point $K_c$ because the background curvature of the system introduces a particular length scale, which suppresses critical fluctuations induced in the planer lattice case\cite{Iharagi2010,Gendiar2014}. [See also $\chi=1024$ in Fig. \ref{fig8}(b)]
Thus, we can straightforwardly extract the correlation length $\xi$ by fitting it in the form of
\begin{align}
\langle \sigma_{N} \sigma_{n} \rangle \sim \exp(-\tilde{\rho}/\xi)\, .
\label{eq_cfit}
\end{align}
The fitting result of the inverse correlation length $1/\xi$ for $h_b=0$ is summarized in Fig. \ref{fig4}.

For the high-temperature region($K<K_c$), it is confirmed that the correlation length $\xi$ is very short and never diverges even at $K=K_c$.
For the low-temperature region($K>K_c$), the deep interior of the system may have a bulk magnetization with spontaneous symmetry breaking.
However, in $K_c < K \lesssim 0.67$, $\langle \sigma_n \rangle=0 $ for $\forall n$ up to $N=10$ due to the free boundary.
Here, we note that the $\langle \sigma_n \rangle=0 $ state in $K_c <K \lesssim 0.67$ is interpreted as a metastable state corresponding to a tachyon mode in the AdS/CFT context.
Then, we have accurately estimated the correlation length with Eq. (\ref{eq_cfit}) for $K < 0.4$ as in Fig. \ref{fig4}.

Next, we calculate the $n$-dependence of a local magnetization along the geodesic line under a tiny boundary magnetic field $h_b = 0.01$.
We then find that $\langle \sigma_n\rangle$ along the geodesic line in the left/right side block of Eq. (\ref{eq_def_rho}) behaves as
\begin{align}
\langle \sigma_{n}\rangle = \langle \sigma_{N-\tilde{\rho}}\rangle \sim \exp(- \tilde{\rho}/\zeta )\, ,
\label{eq_mag_geo}
\end{align}
where $1/\zeta$ denotes the exponential decay rate toward the interior($0<\tilde{\rho} \le N$), or equivalently, $\zeta$ can be termed {\it penetration depth}.
In Fig. \ref{fig4}, we present the fitting result of $1/\zeta$, which decreases as $K$ increases from $K=0.2$.
In particular, one can verify that $1/\zeta =0 $ at $K=K_c$, reflecting a second-order phase transition associated with the bulk magnetization.
For $K>K_c$, the negative $\zeta$ indicates that the disordered state is metastable, corresponding to the tachyon mode, and the magnetic order is actually induced in the deep interior. 
As in the case of $1/\xi$, we can estimate $1/\zeta$ accurately up to $K < 0.4$ with the use of $h_b=0.01$.

Here, we comment on a physical implication of $\zeta$ in the holography context.
For the Bethe lattice Ising model, the effective magnetic field is renormalized by $h^\mathrm{eff}_{N-\tilde{\rho}} \propto (p\tanh K)^{\tilde{\rho}}$ from the boundary to the interior, where $p$ is a branching number for the Bethe lattice model.
For the hyperbolic-lattice model, meanwhile, it is difficult to directly calculate the holographic RG flow for the boundary magnetic field.
Instead, we assume the relation 
\begin{align}
 h^\mathrm{eff}_{N-\tilde{\rho}} \propto \langle \sigma_{N-\tilde{\rho}}\rangle \simeq e^{-\tilde{\rho}/\zeta}, 
 \label{eq_magrg}
\end{align}
which will help us to infer the RG flow associated with the magnetic field when discussing the connection to AdS/CFT later.
Note that this relation is valid for the Bethe lattice model and would be reasonable for the hyperbolic-lattice model, where the correlation length is always very short.

\subsection{boundary spin correlation function}

\begin{figure}[tb]
\begin{center}
\includegraphics[width=7.1cm]{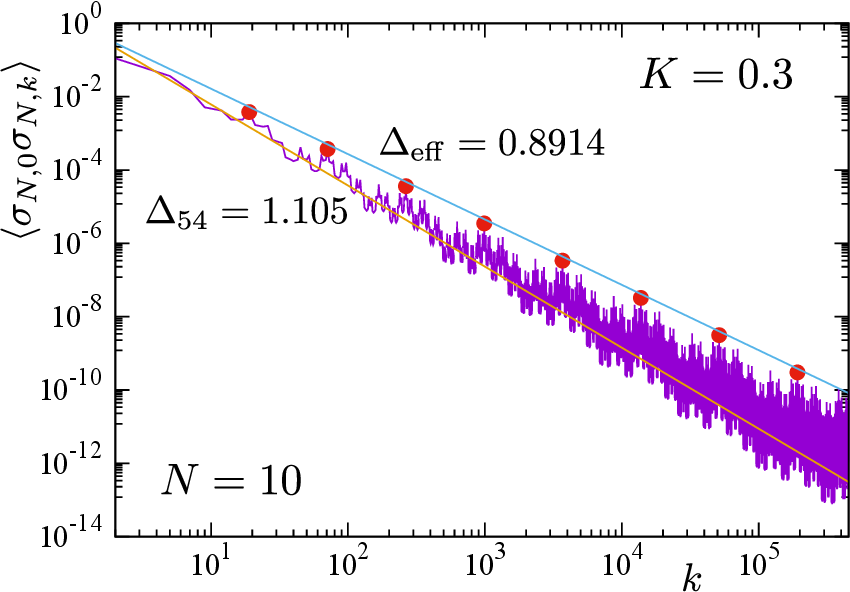}~~
\end{center}
\caption{The boundary spin correlation function for $N=10$ with $K=0.3$. 
The solid line indicates the power law decay of $k^{-2\Delta}$ with $\Delta_{54}=1.105$ and $\Delta_\mathrm{eff} = 0.8914$, which are respectively evaluated by Eqs. (\ref{eq_delta_geo_pq}) and (\ref{eq_delta_geo_tree}).}
\label{fig5}
\end{figure}

In Fig. \ref{fig5}, we show a log-scale plot of the boundary spin correlation function, $\langle\sigma_{N,0}\sigma_{N,k}\rangle$, for the $\{5,4\}$ lattice model with $N=10$ for $K=0.3$(disorder phase).
In the figure, the horizontal axis represents the spin node index $k$ along the circumference direction, where we do not take into account the precise geometric situation of each node, such as its slight deviation from the circle of the radius $\rho=N(=$const) at the level of the hyperbolic space.
Also, we have confirmed that the $N$ dependence of numerical results is almost negligible for $ k < 1.2 \times 10^5$, implying that $N=8$ is sufficient to observe the asymptotic behavior of the boundary correlation function.

Let us precisely analyze the correlation function in Fig. \ref{fig5}, exhibiting power-law decay with quasi-periodic oscillation. 
First, we note that the quasi-periodic oscillation is a general feature of the hyperbolic tiling because the arrangement of the $P$ and $A$ along the circumference direction has quasi-periodicity. 
Such quasi-periodicity for the boundary system is also discussed in terms of the matchgate tensor network\cite{Jahn2022} and the holographic inflation rule\cite{Basteiro2022,Basteiro2023}.
Then, the resulting stable power-law behaviors of the correlation function suggests that the quasi-periodic oscillation originates from the microscopic details around each node and is irrelevant in analyzing their scaling dimensions.

In general, the scaling dimension for the spin-spin correlation function at the outer edge boundary is described by, 
\begin{align}
\langle\sigma_{N,0}\sigma_{N,k}\rangle \sim k^{-2\Delta}\, .
\label{eq_corr_pow}
\end{align}
The essential feature in Fig. \ref{fig5} is that the correlation function can be characterized by two possible scaling dimensions, $\Delta_{54}$ or $\Delta_\mathrm{eff}$, corresponding to the dominant center of the oscillating correlation function and the regular spikes of its upper bound highlighted by red symbols, respectively.
In order to analyze the origin of these scaling dimensions, let us recall the background hyperbolic geometries summarized in Sec. \ref{subsec_geometry};
The distance along the circumference direction is scaled with the number of nodes on the outer edge circle [See Eqs. (\ref{eq_rn_pq}) and (\ref{eq_bnumL})], for which we obtained the two possible radii $L_{54}=0.9422$ and $L_\mathrm{eff}=1,13..$.
On the other hand, the boundary correlation function can also be represented as $\langle\sigma_{N,0}\sigma_{N,k}\rangle  \sim \exp(-\rho/\xi)$, using the geodesic line connecting the two boundary nodes.
In the background hyperbolic plane, the corresponding hyperbolic distance $\rho$ is explicitly given by $ \rho \simeq  2 L \log k$ with $L=L_{54}$ or $L_\mathrm{eff}$ for $k\ll \mathcal{N}_\mathrm{b}$ [See Eqs. (\ref{eq_geoleng_pdisk}) and (\ref{eq_geoleng_puhalf})].
Combining the bulk correlation length $\xi$ for Eq. (\ref{eq_cfit}) and these radii $L$,  we can thus obtain the scaling dimensions for Eq. (\ref{eq_corr_pow}) as
\begin{align}
\Delta_{54} \equiv  L_{54}/\xi
\label{eq_delta_geo_pq}
\end{align}
or
\begin{align}
\Delta_\mathrm{eff} \equiv  L_\mathrm{eff}/\xi \, .
\label{eq_delta_geo_tree}
\end{align}

In Fig. \ref{fig5}, the blue and orange lines illustrate slopes with the scaling dimensions of $\Delta_{54}$ and $\Delta_\mathrm{eff}$, respectively.
We then find that $\Delta_{54}$ explains the dominant center line of the quasi-periodically oscillating correlation function.
On the other hand, the slope with $\Delta_\mathrm{eff}$, which is more relevant than $\Delta_{54}$,  captures a series of regular peak structures turning up in the highly oscillating behavior of the correlation function, which are illustrated as red symbols in Fig. \ref{fig5}. 
A precise analysis of numerical data has revealed the node indices for these peaks to be
\begin{align}
k_\mathrm{peak}= 19, 71, 265, 989, 3691, 13775, 51409, 191861 \cdots \, .
\label{eq_peak}
\end{align}
Interestingly, this sequence corresponds to 
\begin{align}
\mathcal{N}_A + \mathcal{N}_P -2 \, ,
\label{eq_numPA}
\end{align}
in Table 1, indicating that the spin-spin correlation is relatively enhanced just at a distance corresponding to a smaller size CTM consisting of ``$PA$'' with $n(<N)$ embedded in the full-size CTM with $N$.
Here, note that ``$-2$" in Eq. (\ref{eq_numPA}) originates from the node index counted from $k=0$ and the $P$ and $A$ sharing their edges.
Although the hyperbolic lattice contains the homogeneous loop network structure, the most relevant boundary spin correlation can be described by the boundary nodes connected by a simple network path embedded in the hyperbolic-lattice network.


\subsection{Comparison with a free scalar filed in AdS}

As shown above, we have two possible interpretations for the background geometries behind the boundary spin correlation of the hyperbolic lattice model.
In this subsection, we compare the CTMRG results to a scalar field $\phi$ with a mass $m$ in the AdS$_{d+1}$, where $d$ is the dimension of a boundary CFT.
In the context of AdS/CFT, the scaling property of a free scalar field near the AdS boundary($z \to 0$) is described by
\begin{equation}
\phi(z) \sim A z^{\Delta_-} + B z^{\Delta_+} \, , 
\label{eq_adscft}
\end{equation}
with 
\begin{align}
\Delta_+ + \Delta_- = d \, ,
\label{eq_sum_scaling}
\end{align} 
where $A$ denotes a boundary field coupled to the boundary CFT, and $B$ represents a response of the operator against a perturbation with respect to $A$.
For the Bethe lattice model, the above scalar field relation was established by the correspondence: $d=1$, $ A  \leftrightarrow h_b$, $ B  \leftrightarrow \sigma_N $, and $ \Delta_+ =  - \frac{\log(\tanh K)}{\log p} $, where $p$ is the branching number of the Bethe lattice network\cite{Oku2023}.

\begin{figure}[tb]
\begin{center}
\includegraphics[width=6cm]{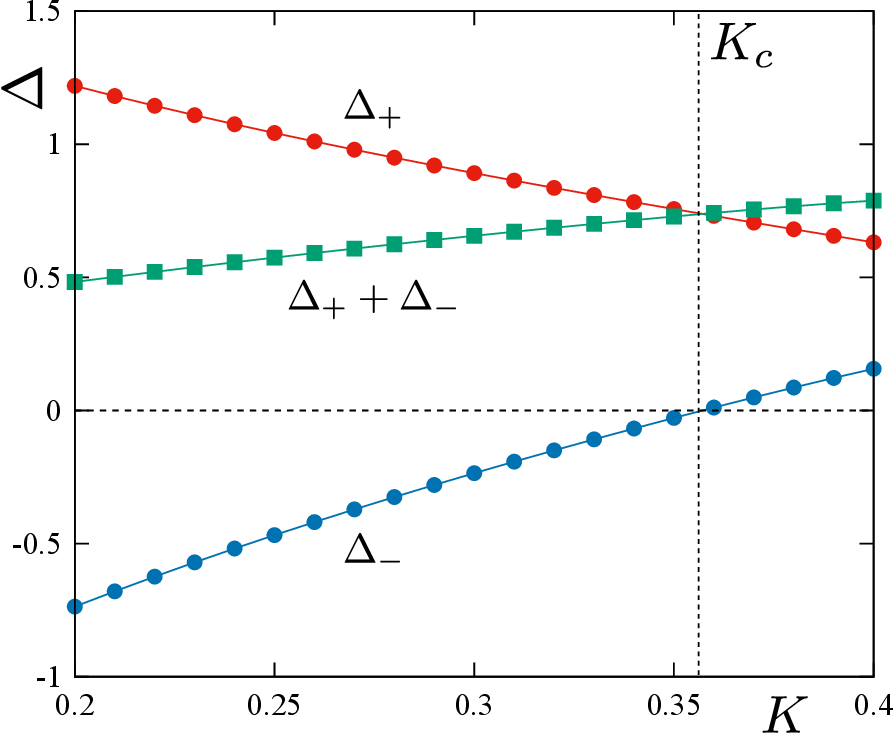}~~
\end{center}
\caption{The scaling dimensions $\Delta_+(=\Delta_\mathrm{eff})$ and  $\Delta_-$ respectively estimated from $1/\xi$ and $1/\zeta$ for $L=L_\mathrm{eff}$. An effective dimension $\tilde{d} \equiv \Delta_+ + \Delta_-$ is also plotted as square symbols.}
\label{fig6}
\end{figure}

From the analysis of the boundary correlation function, we already have the scaling dimension $\Delta_+ = L/\xi$ with $L=L_{54}$ or $L_\mathrm{eff}$ for the $\{5,4\}$ lattice model.
For the hyperbolic-lattice model, however, we do not explicitly have the scaling dimension $\Delta_-$ for $h_b$.
As mentioned in Sec.\ref{sec_3A}, we assume $h^\mathrm{eff}_{N-\tilde{\rho}}  \propto  \langle \sigma_{N-\tilde{\rho}}\rangle$ and read off $\Delta_- = -L/\zeta$ by introducing the radial coordinate $z \simeq e^{-n/L}(\propto e^{\tilde{\rho}/L})$ [See also Eqs. (\ref{eq_poincare_rn}) or (\ref{eq_eff_rn})].
From the two curves of $1/\xi$ and $1/\zeta$ in Fig. \ref{fig4}, we estimate the $K$ dependence of the scaling dimensions, which is depicted in Fig. \ref{fig6}.
In the figure, we also plot $\tilde{d}\equiv \Delta_+ + \Delta_-$ with $L=L_\mathrm{eff}$, indicating the violation of the tree-tensor-network relation (\ref{eq_sum_scaling}) with $d=1$.
This result implies that the loop network structure yields a nontrivial correlation effect on the scaling dimensions.
More systematic analyses for various hyperbolic-lattice models are necessary.
It is also interesting to reexamine the validity of Eq. (\ref{eq_magrg}), which is reminiscent of the free scalar field theory where the roles of $A$ and $B$ can be exchanged for $0 \le m^2L^2+ d^2/4 \le 1$ \cite{Klebanov1999}.

\section{Crossover behavior in the bulk ordered phase}

\begin{figure}[bt]
\begin{center}
\includegraphics[width=7.5cm]{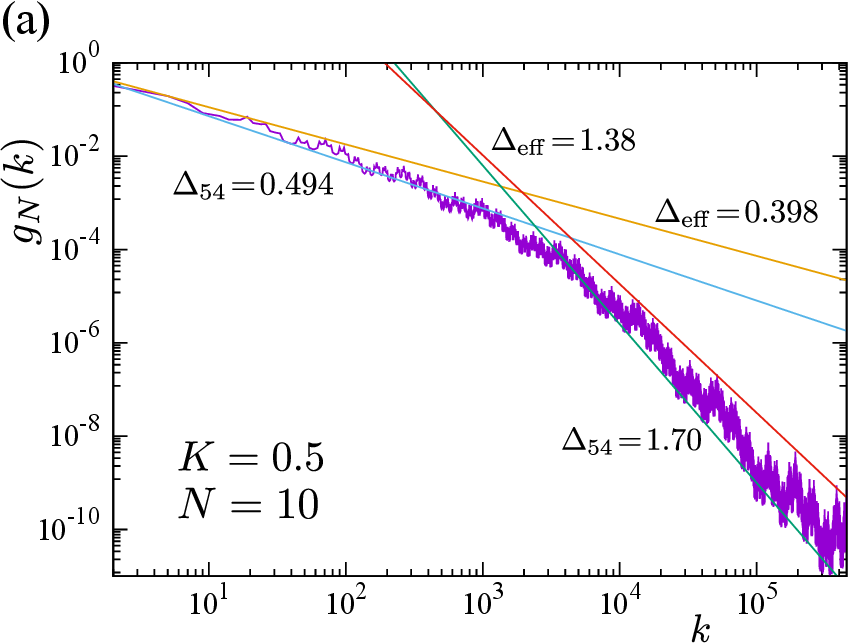}~~
\includegraphics[width=7.5cm]{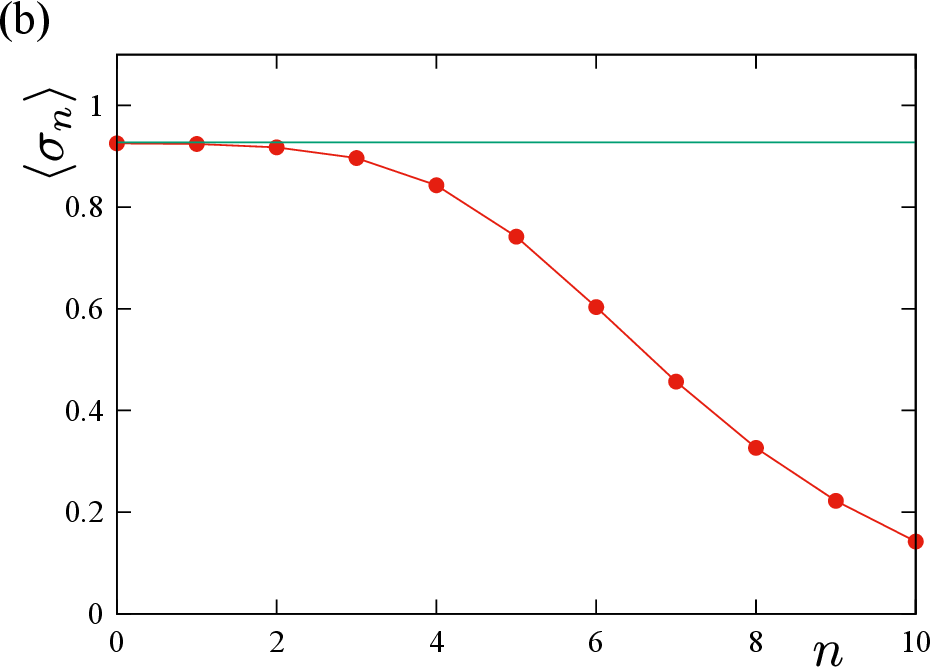}~~\\
\end{center}
\caption{(a) The boundary spin correlation function $g_N(k)$ for $N=10$ with $K=0.5$.
The crossover behavior is observed around $k_\mathrm{cross} \simeq \mathcal{O}(10^3)$.
The slopes with $\Delta_{54}$ and $\Delta_\mathrm{eff}$ are estimated with the use of the bulk correlation length for the disorder fixed point and the bulk ferromagnetic fixed point. 
(b) The generation-index dependence of local magnetizations along the geodesic line through the center node$(n=0)$.
The green horizontal line indicates the bulk expectation value of the local spin, $\langle\sigma\rangle = 0.92735 \cdots$.}
\label{fig7}
\end{figure}

In the bulk-ordered phase, i.e., $K>K_c$, the interior of a sufficiently large hyperbolic lattice has a bulk magnetic order.
As in the Bethe lattice model, one can expect a crossover behavior of a boundary spin correlation function, depending on the depth of the corresponding correlation path.
In order to demonstrate such a crossover for the  $\{5,4\}$ lattice model,  we examine the $N=10$ system at $K=0.5$ in a tiny boundary magnetic field($h_b=0.01$) as a typical example.
Then, we should note that $h_b \ne 0$ induces a small but finite  $\langle \sigma_{N,k} \rangle$ on the outer edge boundary, accompanying quasi-periodic oscillation in $0.079 \lesssim \langle \sigma_{N,k} \rangle \lesssim 0.142$.
By subtracting the background oscillation of $\langle \sigma_{N,k} \rangle$, we here define the boundary spin correlation function as
\begin{align}
g_N(k) \equiv \langle \sigma_{N,0} \sigma_{N,k} \rangle - \langle \sigma_{N,0} \rangle  \langle \sigma_{N,k} \rangle\, .
\label{eq_corr_lt}
\end{align}
Figure \ref{fig7}(a) shows the numerically exact result of $g_N(k)$ obtained by CTMRG computations with no truncation of the bond dimension, where we can confirm a clear crossover behavior of the power-law decay around $k\sim k_\mathrm{cross}\simeq\mathcal{O}(10^3)$.

In order to discuss the crossover behavior from the holographic RG point of view, we have also calculated the local magnetization  $\langle \sigma_n \rangle$ along the geodesic line between the center node and the outer boundary, as depicted in Fig. \ref{fig7}(b).
The essential point is that the magnetization increases from the boundary($n=10$) to the center($n=0$), suggesting that the holographic RG flow of $h^\mathrm{eff}_{N-\tilde{\rho}}$ near the boundary could be characterized by a negative penetration depth governed by the unstable disorder fixed point of $h_b=0$.
Moreover, the bulk correlation length under no magnetic field is estimated as $1/\xi=0.5242\cdots$, which leads us to the scaling dimensions  $\Delta_{54}=0.494$ and $\Delta_\mathrm{eff}= 0.398 $ with the help of Eqs. (\ref{eq_delta_geo_pq}) and (\ref{eq_delta_geo_tree}).
In Fig. \ref{fig7}(a), the slopes corresponding to $\Delta_{54}$ and $\Delta_\mathrm{eff}$ are represented as blue and orange lines, respectively, which are consistent with $g_N(k)$ in the range of $k< k_\mathrm{cross}$.

For $k> k_\mathrm{cross}$, the slope of $g_N(k)$ crossovers to the power characterized by the bulk ferromagnetic fixed point. 
However, unlike the Bethe lattice model, the hyperbolic-lattice model does not have an analytic solution for the fixed point.
Thus, we numerically estimated the correlation length as $1/\xi \simeq 1.81$ by fitting the correlation function in the bulk interior region. (Details of the fitting are not presented here.)
By utilizing this value, we obtain $\Delta_{54}=1.70$ and $\Delta_\mathrm{eff}=1.38$, which respectively correspond to the green and red lines in Fig. \ref{fig7}(a).
Then, the slope of $\Delta_\mathrm{eff}=1.38$ reproduces to the upper edge of the highly oscillating behavior of $g_N(k)$ in $k>k_\mathrm{cross}$.

\section{Cutoff effect of the bond dimension and renormalization group}

\begin{figure}[tb]
\begin{center}
\includegraphics[width=7.2cm]{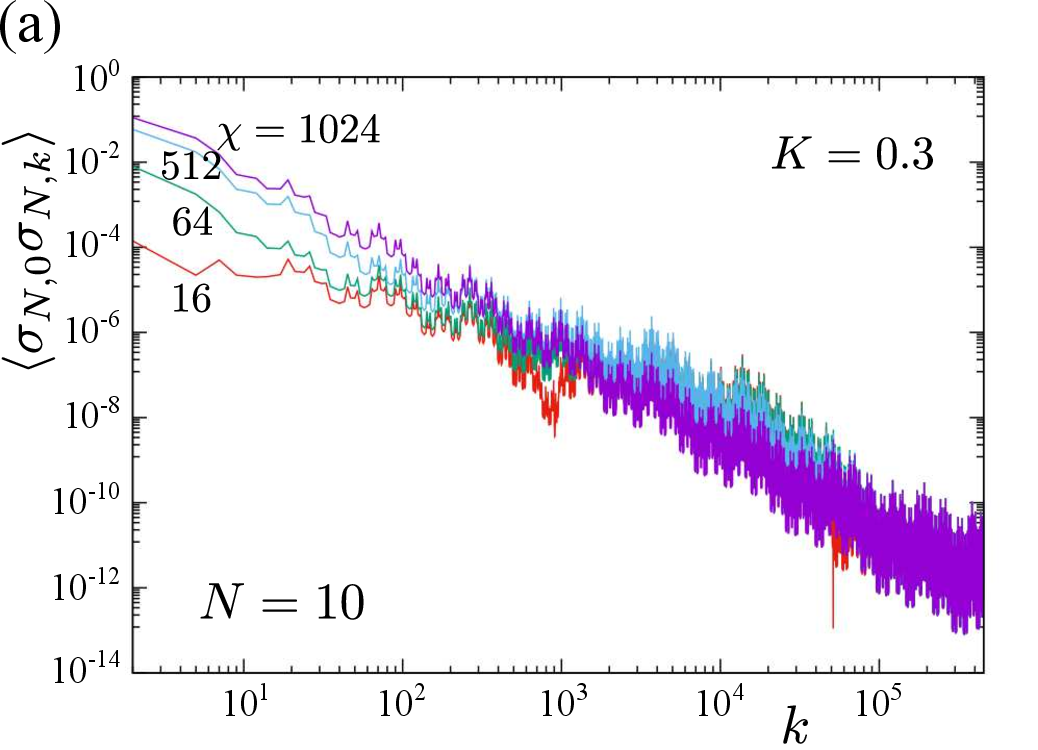}~~
\includegraphics[width=7.1cm]{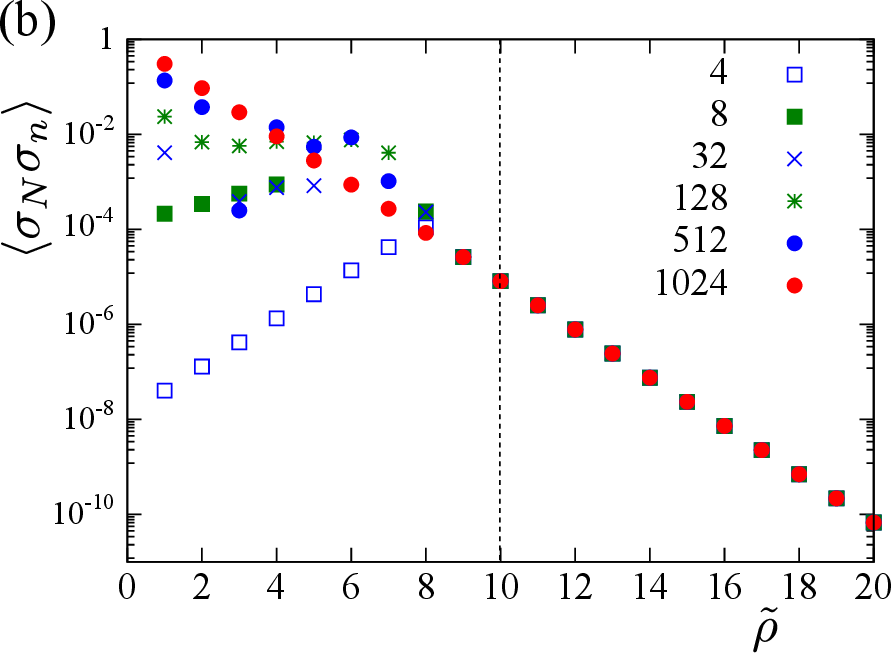}~~
\end{center}
\caption{(a) Cutoff dependence of boundary spin correlation functions for $N=10$ with $K=0.3$. 
The curves indicate $\chi=1024$, $512$, $64$ and $16$ from top to bottom.
(b) Cutoff dependence of bulk spin correlation functions along the geodesic line for $N=10$ with $K=0.3$, where $\tilde{\rho}$ represents the distance of nodes from the boundary node ($\tilde{\rho}=0$) in the left block to its antinodal point ($\tilde{\rho}=2N$) in the right block through the center node ($\tilde{\rho}=N$).
The slope with $\chi=1024$ indicates the exact correlation function with no cutoff.
As $\chi$ decreases, the short-range correlation within the left block rapidly loses accuracy, but the long-range correlation between the left and right blocks maintains the accuracy.}

\label{fig8}
\end{figure}

The CTMRG calculations so far have been done with no cutoff of the bond dimension, and thus, the resulting spin correlation functions are numerically exact.
In the context of holographic tensor networks, nevertheless, it is intrinsically interesting how the cutoff affects the accuracy of the long-distance behavior of boundary spin correlation functions. 
Figure \ref{fig8}(a) shows the cutoff dependence of boundary spin correlation functions for $K=0.3$ with $N=10$.
Note that $\chi=1024$ corresponds to the exact one in Fig. \ref{fig5}.
As $\chi$ decreases, the short-range correlation loses its accuracy, while the long-distance behavior ($k \gtrsim 10^5$) overlaps with the exact one even for $\chi=16$.
This implies that the CTMRG computation for the hyperbolic-lattice model selectively maintains the edge-to-edge correlation (or entanglement) in the renormalized CTM. 
In contrast, the information of short-range correlations is rapidly discarded with repeating RG transformations.
Thus, the above situation of CTMRG concerning the boundary spin correlation function seems consistent with the Kaddanoff-Wilson-type real-space RG, where RG transformations gradually extract longer-scale physics by discarding information of UV scales\cite{WilsonRG,Kadanoff2014}.

In general, such a standard tensor network method as density-matrix renormalization group (DMRG) is suitable for describing the short-range correlation, and critical entanglements are rather hard to represent. 
The CTMRG is equivalent to a matrix product state(MPS) for the planar lattice model\cite{Baxter1968,JPSJ2022}, where the cutoff effect would spoil the quality of longer-distance correlations.
What is the key trick for realizing a proper real-space-RG-like property in the CTMRG for the hyperbolic lattice?
In order to address this question, it is instructive to reconsider the DMRG for a linear chain\cite{White1993};
 it is well established that the edge-to-edge correlation for a finite-length chain can be accurately calculated even for critical cases\cite{Hikihara1998}, implying that a correlation function across the canonical center of the MPS for a finite length can be described well with a finite bond dimension. 
In contrast, a correlation function limited in the left or right block rapidly loses its accuracy.

The essential point in the CTMRG for the hyperbolic lattice is that the geodesic line through the center node corresponds to the chain direction in the DMRG.
In Fig. \ref{fig1}, we consider the geodesic line from the left edge to the antinodal point defined for Eq. (\ref{eq_cfit}) again.
The lower half of the disk in Fig. \ref{fig1}, which is nothing but the product of two CTMs, can then be viewed as a ground-state wavefunction in the DMRG context.
In Fig. \ref{fig8}(b), we show the spin correlation function $\langle \sigma_N \sigma_{n} \rangle $ for $N=10$ along the path, where $\sigma_N$ is located at the left edge and $\sigma_{n}$ moves into the right side. 
The distance $\tilde{\rho}$ between $\sigma_N$ and $\sigma_{n}$ was given by Eq. (\ref{eq_def_rho}).

As demonstrated in Fig. \ref{fig8}(b),  the correlation functions across the center spin of the CTM are actually described well even if the cutoff is introduced, but those for $\tilde{\rho}=1 \sim 8$ rapidly loses accuracy.
In the context of the CTM-based entanglement,  another important observation for the hyperbolic-lattice model is that the decay of the CTM spectrum is very rapid even at the bulk critical point $K_c$, which is consistent with the very short correlation length at $K_c$ in the bulk region (See $\xi$ in Fig. \ref{fig4}).
Thus, a few dominant eigenvalues of the CTM is enough to represent the particular long-range entanglement across the center node. 
Then, the hyperbolic-lattice geometry transmits the high accuracy along the CTM edges into those of the boundary spins near the two edges.
As a result, the CTMRG nature incorporated with the hyperbolic-lattice geometry realizes a selective RG transformation of long-distance correlations.
On the other hand, a massive superposition of the tiny eigenvalue eigenvectors of the CTM is necessary for reproducing the short-range correlation for the outer edge within the left/right block.

From the holographic entanglement point of view, verifying the Ryu-Takayanagi formula for a boundary state defined on the outer edge of hyperbolic-lattice networks is highly desired.
However, we should note that if the hyperbolic lattice Ising model corresponds to a counterpart system on the AdS side, as in the present case, an appropriate definition of entanglement entropy at the level of the tensor network framework remains a nontrivial problem.
On the other hand, if the boundary state of the hyperbolic lattice system is regarded as a wavefunction for a boundary CFT, it may be possible to use the MERA-type algorithms\cite{MERA2007,TNR2015} for computing the corresponding entanglement entropy.

\section{Summary and discussions}

We have precisely investigated the boundary spin correlation function for the hyperbolic-lattice Ising model and its relation to the bulk correlation length along the geodesic line.
The CTMRG computation for the $\{5,4\}$ lattice model has demonstrated that in the high-temperature (disorder) region, the boundary correlation function exhibits power-law decay with quasi-periodic oscillation, while the bulk correlation function always exhibits exponential decay. 
Taking account of the geometric relation of the distance along the outer edge boundary and the bulk correlation path based on the geodesic line, we have found that the most relevant part of the regular peak-like structure in the oscillating correlation function can be characterized by the scaling dimension $\Delta_\mathrm{eff}$ originating from the radius $L_\mathrm{eff}$ for the effective Poincar\'e upper-half based on the exponential inflation of boundary nodes.
On the other hand, we have also revealed that the dominant center line in the quasi-periodically oscillating behavior is explained well by the scaling dimension $\Delta_{54}$ associated with the radius $L_{54}$ for the Poincar\'e unit disk for the $\{5,4\}$ tiling network.
In the low-temperature (bulk ferromagnetic order) region, moreover, we have demonstrated the crossover behavior of the boundary correlation function, depending on the depth of the corresponding correlation path.
Then, the scaling dimensions for the two crossover regime can be also well described by the bulk correlation lengths for the unstable disorder and bulk ferromagnetic fixed points.

These behaviors of the boundary spin correlation functions are basically consistent with the previous result for the Bethe lattice model, a typical example of the tree tensor network. 
However, the correlation due to the loop network structure nontrivially affects the scaling dimensions at the quantitative level.
As in Fig. \ref{fig6}, for instance,  a naive free scalar field relation (\ref{eq_sum_scaling}) with $d=1$ does not hold for the hyperbolic lattice.
How can the loop network effect be translated into AdS/CFT language beyond the tree tensor network?
This question may provide a novel direction of research to gain a deeper understanding of AdS/CFT. 
Then, it may be interesting to reconsider the definition of length scales in hyperbolic-lattice networks beyond the pentagon unit used in this paper.
Also, the validity of Eq.(\ref{eq_magrg}) should be carefully examined.

Here, we should note that the CTMRG calculation suggests a surface transition-like behavior accompanying the spontaneous edge magnetization around $K \simeq 0.67$.
A similar surface spin phase transition is also reported for a hierarchical pentagon network model based on tensor network calculations.\cite{Oshima2024}
In the AdS/CFT context, such an interaction-induced boundary phase transition might cause vacuum instability for the boundary CFT, which may be reminiscent of the Breitenlohner-Freedman instability\cite{BF1982,Basteiro2023BF}.
In addition, the hierarchical pentagon network model, which also exhibits the power-law decay of the boundary spin correlation function, has the same tiling structure as the efficient optimization of path integrals\cite{Caputa2017}.
A detailed analysis of hierarchical models and associated boundary transitions is an interesting future issue.

From the tensor network point of view, we have pointed out that the CTMRG combined with the hyperbolic lattice maintains the particular long-range correlation for the boundary spins near the CTM edges.
Although this behavior is similar to the Kadanoff-Wilson type real-space RG\cite{WilsonRG,Kadanoff2014}, the mechanism of such a selective RG transformation in CTMRG can be explained in analogy with the MPS for the linear chain.
In this sense, it is intriguing to elucidate to what extent the MPS is able to resolve the entanglement structures under the hyperbolic geometry.
However, we should also note that tensor networks such as CTMRG are insufficient for the holographic version of the entanglement entropy associated with outer edge boundary spins in the hyperbolic lattice Ising model.

Recently, a circuit quantum electrodynamics experiment realized a hyperbolic-lattice system.\cite{Kollar2019, Lenggenhager2022}
We hope that the current result stimulates comprehensive research of tensor networks associated with the holographic description of critical correlation/entanglements.



\section*{acknowledgments}
The authors thank T. Takayanagi, A. Gendiar and R. Krcmar for valuable discussion.
This work is supported by Grants-in-Aid for Transformative Research Areas "The Natural Laws of Extreme Universe---A New Paradigm for Spacetime and Matter from Quantum Information" (KAKENHI Grant No. JP21H05182 and No. JP21H05191) from MEXT of Japan.
It is also supported by KAKENHI Grant No. JP21K03403 from JSPS and by the COE research grant in computational science from Hyogo Prefecture and Kobe City through Foundation for Computational Science.


\appendix

\section{Unit disk representations for the hyperbolic plane}
\label{appendix_a}

We briefly summarize the basic properties of the hyperbolic geometry required for discussions in the main text below.

\subsection{Poincar\'e unit disk}
The metric for Poincar\'e unit disk for the hyperbolic plane is given by
\begin{align}
ds^2 = L^2 \frac{4(du^2+dv^2)}{\left(1-(u^2+v^2)\right)^2}
\label{eq_poincare_unitdisk}
\end{align}
for $0 \le u^2+v^2<1$ and $ 0 \le \arctan v/u <2\pi$.
Its scalar curvature and sectional(Gaussian) curvature are  
\begin{align}
R= -\frac{2}{L^2}\, , \quad K= -\frac{1}{L^2}\, .
\end{align}

Let $u=0, v=0$ be the origin and write $u=r\cos\theta$ and $v=r\sin \theta$, which gives
\begin{align}
ds^2 =4 L^2 \frac{dr^2+ (r d\theta)^2}{(1-r^2)^2}\, .
\end{align}
Then, the hyperbolic distance $\rho$ for $0\sim r(<1)$ with a fixed $\theta$ is
\begin{align}
\rho = \int_0^r \frac{2 L dr}{1-r^2} = L\log\left(\frac{1+r}{1-r}\right) \, .
\end{align}
By regarding $\rho$ as a network distance from the center node $n$, the corresponding radius in Poincar\'e unit disk is
\begin{align}
r_n = \tanh\left(\frac{n}{2L} \right) \sim 1- 2e^{-n/L} 
\label{eq_poincare_rn}
\end{align}
for $n \gg L$.
Then, we may write $r_n = 1- z_n$ for $z_n \ll 1$ by introducing $z_n = 2 e^{n/L}$.
Note that $z_n$ represents a distance from the outer circle of the unit disk, which can be related to the the radial coordinate for Eq. (\ref{eq_adscft}).

On the other hand, the circumference of the circle of a radius $\rho$ in the hyperbolic plane is given by
\begin{align}
\oint_{r=\mathrm{const}} ds = \int_{-\pi}^{\pi}  \frac{2 L r d\theta}{1-r^2} = 2\pi L \frac{2r}{1-r^2}
\end{align}
By eliminating parameter $r$, the circumference becomes 
\begin{align}
 2\pi L \sinh \left( \frac{\rho}{L} \right)\, .
\label{eq_hbcircum}
\end{align}
This length is approximately proportional to the number of nodes in the circumference direction of a hyperbolic-lattice model in the main text by equating $\rho$ with the generation index $\rho_n= n$.

\begin{figure}[tb]
\begin{center}
\includegraphics[width=6cm]{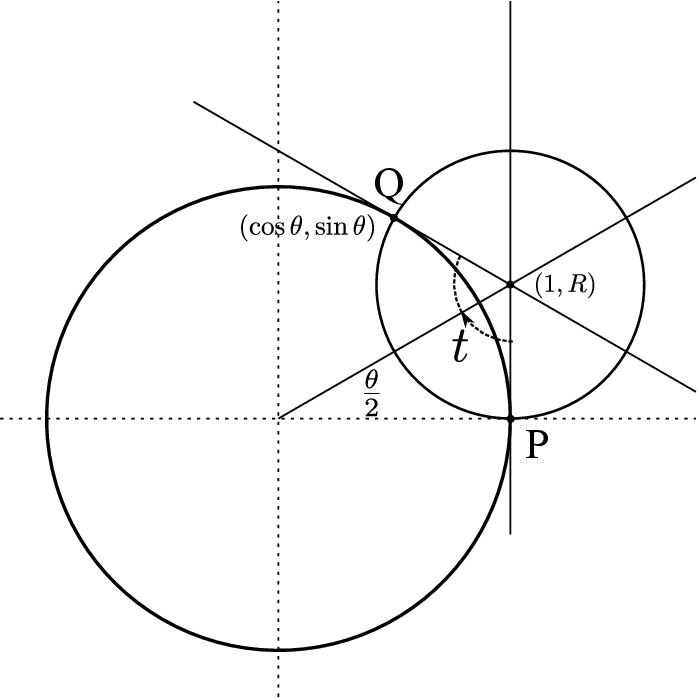}~~
\end{center}
\caption{A geodesic line for the Poincar\'e unit disk. $t$ denotes an angle parameter. }
\label{fig_ap1}
\end{figure}

The geodesic line connecting two boundary points is an arc of the circle orthogonal to the unit circle corresponding to $\rho \to \infty$. 
Figure \ref{fig_ap1} illustrates an example for the case of $\theta=\pi/3$, where $\theta$ is defined as an angle between P$=(1,0)$ and Q$=(\cos\theta, \sin\theta)$.
Then the geodesic circle is $(1-u)^2 + (v-R)^2=R^2$ with $R\equiv \tanh(\theta/2)$, which can be parameterized by
\begin{align}
u&=1-R\sin t \\
v&=R(1-\cos t)
\end{align}
with $ 0 < t < \pi -\theta$.
As depicted in Fig. \ref{fig_ap1}, $t$ corresponds an angle for the geodesic circle from P.
The hyperbolic distance along the geodesic line is 
\begin{align}
\rho(P,Q) &= L \int_\epsilon^{\pi-\theta-\epsilon} \frac{dt}{\sin t -\tan \frac{\theta}{2}(1-\cos t) } \\
&= 2L \log \left( \cot \frac{\epsilon}{2} - \tan \frac{\theta}{2}\right)
\end{align}
where $\epsilon$ is a cutoff angle.
By introducing a cutoff length scale with $a \equiv R \epsilon =  \tan \frac{\theta}{2} \, \epsilon $,  we then obtain
\begin{align}
\rho(P,Q) = 2L \log \left( \tan \frac{\theta}{2} (\frac{2}{a} -1) \right) \sim 2L \log  \frac{2 \tan \frac{\theta}{2}}{a} \, ,
\end{align}
for $a \ll 1$.
By writing the length between P and Q along the unit circle as $x \equiv \theta$, we have
\begin{align}
 \rho(P,Q) \sim  2L \log \left( \frac{2 \tan \frac{x }{2}}{a} \right)\, .
 \label{eq_geoleng_pdisk}
\end{align}
If $a \ll x \ll 1$,  we have $\rho(P,Q) \sim 2L \log \frac{x}{a}$.

\subsection{An effective Poincar\'e upper-half plane}
\label{appendix_a2}

As discussed for Eq. (\ref{eq_bnump}), the number of edge nodes in the hyperbolic lattice increases with the inflation rule $\mathcal{N}_b = \kappa p_\mathrm{eff}^n$ with respect to a generation index $n$, where $\kappa$ is a constant independent of $n$.  
We then define the radius on the unit disk for a generation index $n$ as 
\begin{align}
r_n = 1 - z_n \,,
\label{eq_eff_rn}
\end{align}
with
\begin{align}
z_n  = p_\mathrm{eff}^{-n} =  e^{-n/L_\mathrm{eff}}\, 
\end{align}
in analogy with the Bethe lattice network.\cite{Oku2023}
Assuming the uniform distribution of nodes in the circumference direction, we also introduce
\begin{align}
x_k \propto \frac{2 \pi r_n}{ \kappa p_\mathrm{eff}^n}  k \simeq \frac{2 \pi}{\kappa}  \frac{1-z_n}{z_n}k  
\end{align}
where $\kappa$ is also independent of $k$.
By smearing out the network coordinate $(z_n, x_k)$, we have an effective metric  for $z \ll 1$ 
\begin{align}
ds^2 = L_\mathrm{eff}^2 \frac{dz^2+d\tilde{x}^2}{z^2} \, ,
\label{eq_upperhalf}
\end{align}
where $L_\mathrm{eff} = 1/\log p_\mathrm{eff}$ and $x$ is rescaled by $\tilde{x}\equiv  \frac{1}{2\pi \kappa \log p_\mathrm{eff}}$, as in the case of the Bethe lattice model.
This coordinate corresponds to the effective Poincar\'e upper half with compactification for $\tilde{x}$ due to the periodic boundary condition.
The geodesic length for a path connecting two points, P and Q, on the outer edge boundary is given by 
\begin{align}
\rho(P,Q) \sim 2 L_\mathrm{eff} \log \left(\frac{2 \sin \frac{\tilde{x}}{2} }{a}\right) \, ,
\label{eq_geoleng_puhalf}
\end{align}
where $a$ is a cutoff scale. 
Note that $\tilde{x}$ corresponds to the distance along the circumference of the outer edge circle.

Here, it should be emphasized that Eq. (\ref{eq_upperhalf}) is not directly related to Eq. (\ref{eq_poincare_unitdisk}) by the standard conformal mapping ($w=\frac{z-i}{z+i}$).
As the geodesic line comes into the deep interior of the unit disk, thus, Eq. (\ref{eq_geoleng_puhalf}) deviates from the geodesic length of Eq. (\ref{eq_geoleng_pdisk}) for the Poincar\'e unit disk.
Here, note that the definition of the cutoff scale in Eq. (\ref{eq_geoleng_puhalf}) does not correspond to $a$ in Eq.(\ref{eq_geoleng_puhalf}). 
Also, Eq. (\ref{eq_upperhalf}) is valid only for $z \ll 1$ while Eq. (\ref{eq_poincare_unitdisk}) is able to describe Porencare's unit disk correctly.

\bibliographystyle{ptephy}
\bibliography{penta}

\end{document}